\documentclass[english,american,aps,prl,superscriptaddress,reprint]{revtex4-1}
\usepackage[utf8]{inputenc}
\usepackage{xcolor}
\usepackage{babel}
\usepackage{mathtools}
\usepackage{amsmath}
\usepackage{graphicx}
\usepackage[unicode=true,
 bookmarks=true,bookmarksnumbered=false,bookmarksopen=false,
 breaklinks=false,pdfborder={0 0 1},backref=false,colorlinks=true]
 {hyperref}
\hypersetup{
 pdfborderstyle=,pdfborderstyle={},allcolors=blue}

\makeatletter

\newcommand*\LyXThinSpace{\,\hspace{0pt}}

\usepackage{babel}

\usepackage{bbold}

\usepackage{algpseudocode}

\newcommand{\change}[1]{ {\color{black} #1}}

\DeclareTextSymbolDefault{\textquotedbl}{T1}
\makeatother

\begin{document}
\title{ Easy access to energy fluctuations in
non-equilibrium quantum many-body systems}
\author{Marcela Herrera}
\email{Electronic Address: amherrera@uao.edu.co}

\affiliation{Departamento de Ciencias Naturales, Universidad Autónoma de Occidente,
Cali, Colombia}
\affiliation{Centro de Ciências Naturais e Humanas, Universidade Federal do ABC,
Avenida dos Estados 5001, 09210-580 Santo André, São Paulo, Brazil}
\author{John P. S. Peterson }
\affiliation{Institute for Quantum Computing and Department of Physics and Astronomy,
University of Waterloo, Waterloo N2L 3G1, Ontario, Canada}
\author{Roberto M. Serra}
\email{Electronic Address: serra@ufabc.edu.br}

\affiliation{Centro de Ciências Naturais e Humanas, Universidade Federal do ABC,
Avenida dos Estados 5001, 09210-580 Santo André, São Paulo, Brazil}
\author{Irene D'Amico}
\email{Electronic Address: irene.damico@york.ac.uk }

\affiliation{Department of Physics, University of York, York YO10 5DD, United Kingdom}
\begin{abstract}
We combine theoretical and experimental efforts to propose a method
for studying energy fluctuations, in particular, to obtain  {the related} bi-stochastic
matrix of transition probabilities by means of simple measurements
at the end of a protocol that  drives a  {many-body} quantum
system out-of-equilibrium. This scheme is  {integrated}
with numerical optimizations in order to ensure a proper analysis
of the experimental data, leading to physical probabilities.  The
method is experimentally evaluated employing a two interacting spin-1/2
system in a nuclear magnetic resonance setup. We show how to recover
the transition probabilities using only local measures which enables
an experimental verification of the detailed fluctuation
theorem in a  {many-body} system driven out-of-equilibrium.
\end{abstract}
\maketitle
 Energy fluctuations play a significant  {role on the out-of-equilibrium thermodynamics} of quantum systems \cite{Goold2016,Vinjanampathy2016,Kosloff2013}.
 {They are} inherently related to fluctuation relations \cite{Jarzynski1997,Crooks1999,Kurchan2000,Mukamel2003,Talkner2007b,Esposito2009,Campisi2011,Hanggi2015},
which embraces both thermal and quantum energy fluctuations. In this context
thermal fluctuations are, generally, related with thermal distributions
 {at} the beginning of a driving protocol or  {at} the end of a thermalization
process, while quantum fluctuations are associated to transitions
between eigenstates in a quantum dynamics, depending on how the systems
is driven. \change{Here,} work and heat are described by stochastic
variables with probability distributions \cite{Goold2016,Vinjanampathy2016,Kosloff2013,Hanggi2015}.
The experimental verification or use of quantum fluctuation relations
requires the assessment of both types of energy fluctuations \cite{Camati2018}. 

 Here we introduce a  {powerful} method to experimentally
 {access} energy fluctuations of a many-body system in an out-of-equilibrium
quantum evolution. More specifically we show how to reconstruct  {the} bi-stochastic
matrix of transition probabilities  {$p_{m|n}$} between the initial and final eigenstates
of a driven protocol  {that determines quantum fluctuations. We then} use the matrix to reconstruct the quantum work
probability distribution. It can also be used to reconstruct the statistics
of other quantities such as heat in the absence of work.  {Previous methods}
for this purpose are very demanding when applied to  {many-body} systems since they involve controlled operations, as the interferometric
method proposed in~\citep{Dorner2013,Mazzola2013}.  {In fact, so far,  they have been employed in NMR experiments with one-body (i.e. two level) quantum systems, such as  in Refs.~\citep{Batalhao2014,Batalhao2015,Camati2016,Peterson2018}
or in the} quantum work meter implemented on an ensemble of non-interacting two-level atomic systems~\citep{Cerisola2017}.
 An efficient protocol to experimentally investigate
energy fluctuations in a general out-of-equilibrium many-body system remains an outstanding challenge:
 the aim of this letter is to provide a long stride towards this goal by introducing a fresh approach.  \change{Previously we developed methods inspired by density-functional-theory (DFT~\cite{Jones2015})} \cite{Herrera2017,Herrera2018,Skelt2019}   to support numerical calculations of energy fluctuations for out-of-equilibrium many-body system, a tough problem in itself; here we focus on providing directly an experimental method for measuring these quantities.

The transition  {probabilities among instantaneous eigenstates $p_{m|n}$
 allow} us to access the statistical
properties  of a time-dependent driven system and
it is intimately related to quantities in the out-of-equilibrium thermodynamics
as the work and entropy production.  The work is one
of the key  {quantities} to describe the change of energy when an external
agent acts on the system. The work performed on a quantum system
is not an observable, but instead, it is determined by the way the
process is executed~\citep{Talkner2007,Talkner2016}.  {For} a protocol much faster than the typical interaction
time scale with the \change{environment \cite{note1} }, the mean work performed on or by
a quantum system can be written as  {
$\langle W\rangle=\int P(W) dW$, where the work probability distribution is $P(W)=\sum_{n,m}p_{n}^{0}p_{m|n}\delta\left(W-\mathcal{\epsilon}_{m}^\tau+\mathcal{\epsilon}_{n}^0\right)$,
with} $p_{n}^{0}$ is the probability to find the system in the initial
eigenstate $|n(0)\rangle$, the transition probability $p_{m|n}=\left|\langle m(\tau)|\mathcal{U}_{\tau}|n(0)\rangle\right|^{2}$
is the conditional probability of driving the system to the instantaneous
eigenstate $|m(\tau)\rangle$ (with energy  { $\mathcal{\epsilon}_{m}^\tau$})
at the end of the evolution, given the initial state $|n(0)\rangle$(with
energy  { $\epsilon_{n}^0$}), where $\mathcal{\mathcal{U}_{\tau}}$ is
the time evolution operator. We note that the transition probability
$p_{m|n}$ is the quantity that defines how the driving processes
is performed and it is also a key element in the determination of
the entropy production in the protocol~\citep{Esposito2009,Seifert2012,Batalhao2015}.

Conventionally, we can obtain $p_{m|n}$ by means of  {the
so-called} two-point measurement protocol~\cite{Mukamel2003,Monnai2005,Campisi2011,Talkner2007}.
In this protocol, the system is initially prepared in the thermal
equilibrium state, then the eigenstates of the initial Hamiltonian
$\mathcal{H}_{0}$ are measured, through a non-destructive measurement
represented by the projectors ${\Pi_{n}^{0}=|n(0)\rangle\langle n(0)|}$.
After that, the system is evolved due to the variation of a parameter
of the Hamiltonian, whose evolution is described by the operator $\mathcal{U}_{\tau}$.
Finally, the eigenstates of the final Hamiltonian $\mathcal{H}_{\tau}$
are measured and this measurement is represented by the projectors
${\Pi_{m}^{\tau}=|m(\tau)\rangle\langle m(\tau)|}$. Eventually, from the measurement of the full statistics, it is possible to obtain $p_{m|n}$. The extension
of the two-point measurement protocol to open systems was proposed
in~\textcolor{black}{\citep{Suomela2014}. }Performing such non-destructive
projective measurements with high accuracy in an experiment is a very
difficult task, even in a few-body system.\textcolor{black}{{} Other
theoretical protocols include generalized energy measurements as Gaussian
energy measurements}~\textcolor{black}{\citep{Watanabe2014} and
positive operator valued measurements (POVM)}~\textcolor{black}{\citep{Roncaglia2014}}.
Another  possibility is to use an interferometric
protocol as employed in Refs.~\citep{Dorner2013,Mazzola2013,Batalhao2014,Batalhao2015,Peterson2018}
that requires a good control of each part of the many-body system
and of the interaction between its constituents, which is challenging
to implement experimentally  {even in few-body} systems.  {Here, we propose
an alternative way to obtain the transition probabilities $p_{m|n}$ through
the direct determination of a set of observables, {\it easy to measure} } in
a given \change{context.}

 {\emph{Inversion scheme.}} Our protocol relies
on an inversion scheme to obtain $p_{m|n}$ and is based on similar
ideas to those given in Ref.~\citep{Rohringer2006} in the context
of DFT. Let us consider measuring the mean value of an operator $\mathcal{O}$
at the end of a   {time-dependent }protocol described
by the evolution operator $\mathcal{\mathcal{U}_{\tau}}$ that drives
the system. We assume that initially the system is prepared in a thermal
state $\rho_{0}^{eq}$, for a given \change{thermal energy $k_B T$},
with an initial Hamiltonian $\mathcal{H}_{0}$. The system will evolve
according to $\mathcal{\mathcal{U}_{\tau}}$ and during this process
all allowed transitions between energy levels may occur. At time $\tau$,
corresponding to the end of the evolution, the mean value of a  {suitable}
observable, $\langle\mathcal{O}\rangle$, is measured.
We can decompose  { this} in terms of the instantaneous Hamiltonian
 energy basis as $\langle\mathcal{O}(\tau)\rangle=\sum_{m,m',n}\mathcal{O}_{m,m'}p_{n}^{0}T_{m,m'|n},$
where $\mathcal{O}_{m,m'}$ are the operator matrix elements written
on the basis of the final Hamiltonian $\mathcal{H}_{\tau}$, $p_{n}^{0}$
is the initial Gibbs distribution, and $T_{m,m'|n}=\langle m(\tau)|\mathcal{\mathcal{U}_{\tau}}|n(0)\rangle\langle n(0)|\mathcal{\mathcal{U}_{\tau}}^{\dagger}|m'(\tau)\rangle$
represents the transition elements, with $p_{m|n}=T_{m,m|n}$. It
is important to highlight that the transition elements are \foreignlanguage{english}{\textit{independent}
both from the initial temperature and from the choice of $\mathcal{O}$.
Instead, they only depend on how ('fast' or 'slow') the evolution
of the system is performed. Hence, we may choose an operator $\mathcal{O}$
which is diagonal in the basis of $\mathcal{H}_{\tau}$, or in other
words $\left[O,\mathcal{H}_{\tau}\right]=0$, thus $\langle\mathcal{O}(\tau)\rangle=\sum_{m,n}\mathcal{O}_{m}p_{n}^{0}p_{m|n}$
and we recover the transition probability $p_{m|n}$ between the instantaneous
eigenstates of the initial and final \change{Hamiltonians\cite{note2}}.}

\begin{figure}
\begin{centering}
\includegraphics[width=0.98\columnwidth]{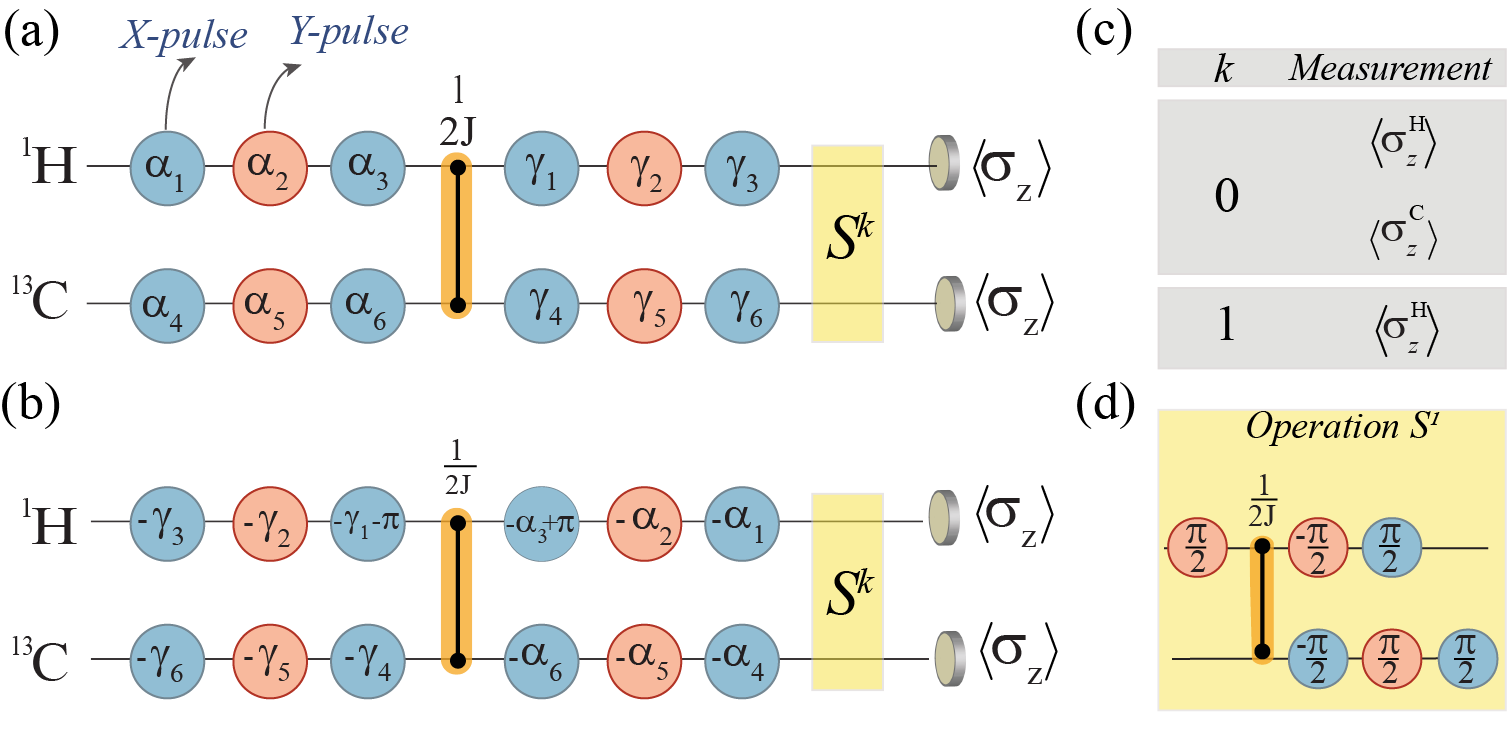}
\par\end{centering}
\caption{\textbf{ {\small{}Pulse sequence for the driven dynamics
and measurement protocols. }} {\small{}In the sketch
of the pulse sequence the blue (red) circles represent transverse
rf pulses in the $x$ ($y$) direction that produce rotations by the
displayed angle. The orange connections represent free evolutions
under the scalar interaction $\mathcal{H}_{J}=h J\sigma_{z}^{H}\sigma_{z}^{C}/4$
(with $J\approx215.1$ Hz) during the time   {$1/(2J)$}. (a) Displays the
sequence to implement the forward protocol. (b) Represents the sequence
to implement the backward version of the  {driving} protocol. (c) Indicates
which local  {measures should} be carried out depending on the value
of $k$. (d) Represents the operation $S^{1}$ ($S^{0}$ is the identity).
The  {values for} $\alpha_{i}$ and $\gamma_{i}$ are displayed in \cite{SuppMat}.}}
\label{fig:Foward and Backward circuit}
\end{figure}

The key idea of our method is to write a system of linear equations
based on the  {mean} value $\langle\mathcal{O}\rangle$,
where the set of variables are given by the elements $p_{m|n}$. For
a system of Hilbert space dimension $d$, we should find $d\times d$
different transition probabilities. However, the square matrix $p_{m|n}$
is a bi-stochastic matrix, implying that, for each $m$ and $n$,
the system is subject to the normalization conditions $\sum_{m}p_{m|n}=1$
and $\sum_{n}p_{m|n}=1$. Our system of linear equations can then
be reduced to $(d-1)\times(d-1)$ elements. Thereby, if we have a
two-level system,  {obtaining the corresponding four transition probabilities with the use of the normalization conditions intakes
finding just one additional equation.} For a system with a higher dimension
we can find $\left(d-1\right)\times\left(d-1\right)$ equations by
using different temperatures and/or measuring additional operators,
so to construct a matrix equation of the form $\mathcal{A}\boldsymbol{x}=\boldsymbol{b}$
as{\small{} }\cite{SuppMat}{\small{}
\begin{equation}
\mathcal{A}=\begin{bmatrix}a_{11}^{1} & a_{12}^{1} & \cdots & a_{d',d'}^{1}\\
a_{11}^{2} & a_{12}^{2} & \cdots & a_{d',d'}^{2}\\
\vdots & \vdots & \ddots & \vdots\\
a_{11}^{d'\times d'} & a_{12}^{d'\times d'} & \cdots & a_{d',d'}^{d'\times d'}
\end{bmatrix}\hspace*{-0.15cm},\;\boldsymbol{x}=\begin{bmatrix}p_{1|1}\\
p_{1|2}\\
\vdots\\
p_{d',d'}
\end{bmatrix}\hspace*{-0.15cm},\;\boldsymbol{b}=\begin{bmatrix}b_{1}\\
b_{2}\\
\vdots\\
b_{d',d'}
\end{bmatrix}\label{eq:general}
\end{equation}
}where $d'=d-1$, \textbf{$\boldsymbol{b}$}  {represents constants}
independent of $p_{m|n}$ and the superscript in  {the constant coefficients}  $a_{mn}^{j}$ represents
the different choices of operators and temperatures. Additionally,
we can also use the symmetries of the system to further reduce the
number of variables. To summarize, we aim to combine different observables
and initial temperatures to get enough linear independent equations
to allow reconstruction of the bi-stochastic matrix $p_{m|n}$. The
number of observables needed depends on the symmetries of the initial
and final Hamiltonians. \change{As the  measurements are done at the final time, they are independent from the energy difference across the transition. In the spirit of DFT, we aim to use as observable
local magnetization or local particle densities.
We present ideas for general applications of the method to large systems in the supplemental material\cite{SuppMat}.}

\begin{figure*}
\begin{centering}
\includegraphics[scale=0.27]{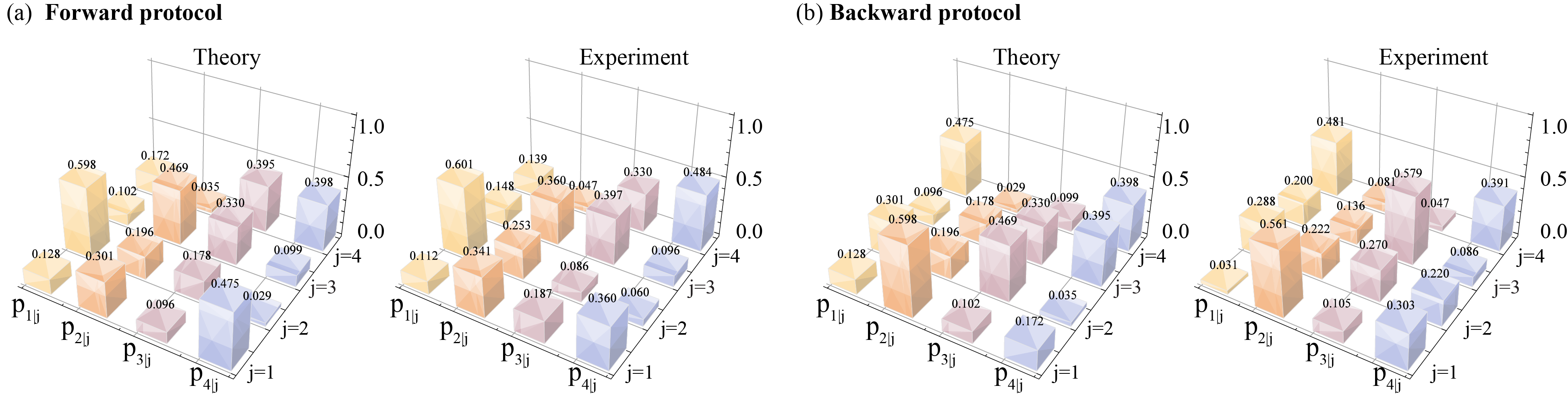}
\par\end{centering}
\caption{\textbf{ {\small{}Transition probabilities.}} {\small{}
Transition probabilities obtained by the inversion scheme for the
(a) forward and  { (b) backward} protocols . For each protocol
we show the comparison between the expected result given by the  {numerical} simulation
of the protocol (theory) and the experimental result. The elements
($p_{i|j}$) of the transition probability matrix are represented
by the  {columns} in the figure and their values are above the  {columns}.}}
\label{fig:FIG2}
\end{figure*}

 {\emph{Experimental implementation.}} In order
to test our proposed scheme for many-body systems let us consider
an out-of-equilibrium evolution for a system of two interacting spins.
 This is significantly more complex than the single-qubit
systems measured in some previous quantum thermodynamics experiments~\citep{Batalhao2014,An2014,Batalhao2015,Peterson2016,Camati2016,Peterson2018},
as the number of transitions $p_{m|n}$ changes from 4 to 16.  {For the experiment,}
 we employed a $^{13}$C-labeled CHCl$_{3}$ liquid sample
and a 500 MHz Varian NMR spectrometer.  The relevant
nuclear spin Hamiltonian for this molecule, in a rotating frame~\cite{SuppMat}, is similar to
the Ising model and can be written as \cite{Oliveira2007} 

\begin{equation}
\mathcal{H}=-\frac{1}{2}h\delta\nu_{\text{H}}\sigma_{z}^{\text{H}}-\frac{1}{2}h\delta\nu_{\text{C}}\sigma_{z}^{\text{C}}+\frac{1}{4}h J\sigma_{z}^{\text{H}}\sigma_{z}^{\text{C}},\label{eq:chloroformham}
\end{equation}
where $\delta\nu_{H}$ and $\delta\nu_{C}$  { are the difference
between the Larmor frequency and the rf-field frequency
for the Hydrogen and the Carbon nuclei, respectively,} and $J\approx215.1$~Hz
is the coupling constant. For this experiment we have chosen $\delta\nu_{\text{H}}=2.0$~kHz
and $\delta\nu_{\text{C}}=4.0$~kHz.  Time-modulated
rf-field pulses in the transverse ($x$ and $y$) direction combined
with longitudinal field gradient pulses are used to prepare initially
thermal states. The rf-field pulses can also be used to drive the
nuclear spins provoking transitions between the eigenstates of the
Hamiltonian.

 For testing our method, we consider a general driven
evolution of the two spins-1/2 system, which is implemented through
a set of rf pulses and free evolutions. At the beginning and at the
end of the evolution, the Hamiltonian of the system will be given
by Eq.~\eqref{eq:chloroformham} defining the reference energy levels.
As demonstrated in Ref.~\citep{Vatan2004} a general two-qubit evolution
can be realized by a circuit consisting of 12 elementary one-qubit
gates and 2 CNOT gates. Since our goal is to  {perform
a proof-of-principle experiment}, we are not concerned in knowing
what is the  {exact time-dependency} in the Hamiltonian that implements such
evolution. What we want, is to implement an evolution that produces
as many as possible non-zero transition probabilities between the
initial and final energy eigenstates, in order to produce a non-trivial
work probability distribution. To this aim, we can consider just a
subclass of the general  {evolutions} proposed in Ref.~\citep{Vatan2004}.
 {To test the detailed quantum fluctuation relation~\citep{Talkner2007b}, we implement forward ($\mathcal{U}_{F}$) and backwards ($\mathcal{U}_{B}$)  evolutions,
by applying the unitary operations depicted in the pulse sequences
of Fig.~\ref{fig:Foward and Backward circuit}, (a) and (b), respectively.}    {For $\mathcal{U}_{B}$ to be the time reversal evolution of $\mathcal{U}_{F}$,}
we have to apply all the pulses in the inverse  {order} starting
with the angles $\gamma_{i}$ and finishing with the angles $\alpha_{i}$.
 {Also, $\gamma_{i}\to -\gamma_{i}$ and $\alpha_{i}\to -\alpha_{i}$ to }satisfy
$\mathcal{U}_{B}=\mathcal{U}_{F}^{\dagger}$.
 { Finally, by an appropriate} choice of the angles $\alpha_{i}$ and $\gamma_{i}$ we can obtain
an out-of-equilibrium evolution that produces several transitions
between the initial and final eigenstates.
\begin{figure*}
\begin{centering}
\includegraphics[scale=0.29]{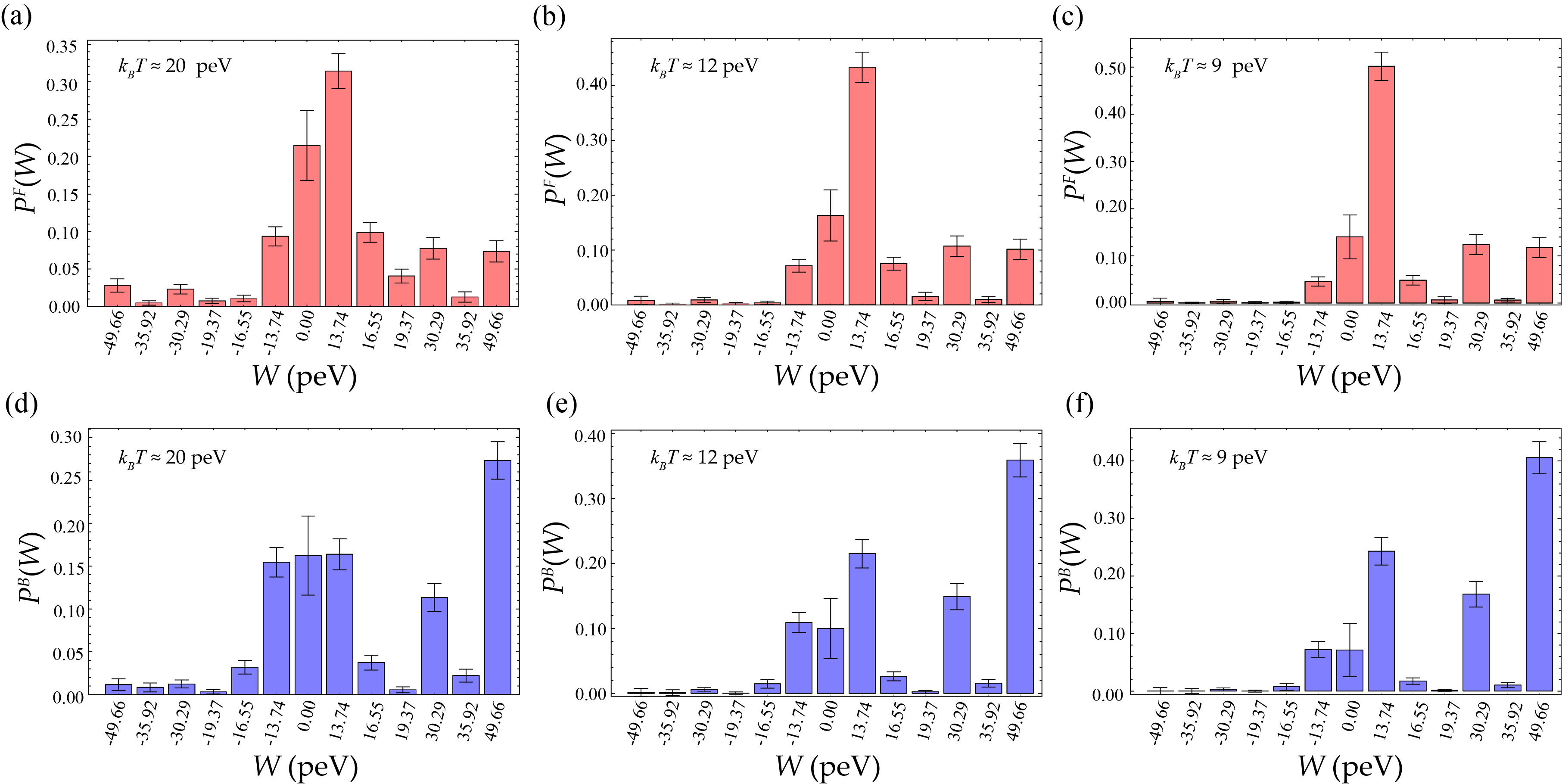}
\par\end{centering}
\caption{ {\bf  Experimental result for the the work distribution}.
Work distributions for the   {forward (a,b,c)} and   {backward (d,e,f)} protocols
for the three different initial spin temperatures (a, d) \change{$k_B T=20\pm3$~$\text{peV}$},  (b, e) \change{$k_B T= 12\pm2$~$\text{peV}$} 
and  (c, f) \change{$k_B T= 9\pm2$~$\text{peV}$}. }
\label{fig:FIG3}
\end{figure*}

 Since we are dealing with a system of dimension 4
(two spin 1/2),  {there will be a } total 16 transition probabilities
that describe the work distribution in the driving protocol. By using
the bi-stochastic properties of the  transition matrix
$p_{m|n}$, our problem reduces to find 9 transitions probabilities
only. Hence, in order to construct the system of equations~\eqref{eq:general},
we can measure three observables -- chosen to be the  {longitudinal
magnetizations} $\sigma_{z}^{\text{H}}$, $\sigma_{z}^{\text{C}}$,
and  {the correlation function} $\sigma_{z}^{\text{H}}\sigma_{z}^{\text{C}}$
-- at three different effective spin temperatures, \change{$k_B T_{1}= 20\pm3$~$\text{peV}$},
 \change{$k_B T_{2}=12\pm2$~$\text{peV}$}, and  \change{$k_B T_{3}= 9\pm2$~$\text{peV}$}
(for more details related to the initial thermal states see~\citep{SuppMat}).
 These  {chosen observables commute} with the Hamiltonian~\eqref{eq:chloroformham},
 {which simplify the  {numerical analysis} in the inversion
scheme}. To obtain $\langle\sigma_{z}^{\text{H}}\rangle$ and $\langle\sigma_{z}^{\text{C}}\rangle$
in the NMR experiment,  {we apply a $\pi/2$ rotation
in the $y$-direction and measure the traverse magnetization, which
is the natural observable in NMR  $(k=0$ in Fig.~\ref{fig:Foward and Backward circuit}).}
For obtaining the correlation function $\langle\sigma_{z}^{\text{H}}\sigma_{z}^{\text{C}}\rangle$,
we consider $k=1$, with the operation $S^{1}$ being just a CNOT
gate written in terms of the rf pulses. Then at the end of the protocol
we have to measure the magnetization in the $z$-direction of the
Hydrogen $\langle\sigma_{z}^{H}\rangle$, as indicated in the grey
table of Fig.~\ref{fig:Foward and Backward circuit}(c). Therefore,
at the end of the day, we are measuring the local magnetizations for
obtaining all the observables for both the forward and backward processes.
 {We note that, depending on the experimental context
different observable can be chosen.}  {In the case
of an} electronic system what would be measured is the  {local} particle density~\citep{Rohringer2006}.

Due to experimental noise and imperfections, the direct solution from
the system of equations could lead to a non physical transition probability
matrix. In other words, we could obtain negative values for $p_{m|n}$'s
or values greater than one. A similar problem happens in the experimental
implementation of quantum state tomography (QST)~\cite{Chuang1997,Nielsen2011,Banaszek1999,James2001},
and its source is the fact that different experimental arrangements
(measurement apparatus) are needed to measure  different
observables. Each experimental arrangement carries a particular noise
that may lead to nonphysical values. In order to deal with this kind
of experimental errors,  {the Maximum Likelihood
Estimation (MLE)\citep{Banaszek1999,James2001} method is commonly} used for QST.  This
method requires numerical optimization to generate a  {definite} positive
density matrix given a set of  {experimental} data  {from} quantum state tomography.
 To implement the MLE, a likelihood function is introduced
that allows to determine how close the physically estimated density
matrix fits the experimental data. Motivated by the successes in
quantum state tomography, we propose   {a MLE method adapted}
to our problem for obtaining physical transition probabilities for
experimental data with noise and imperfections \citep{SuppMat}.

\begin{figure*}
\begin{centering}
\includegraphics[scale=0.23]{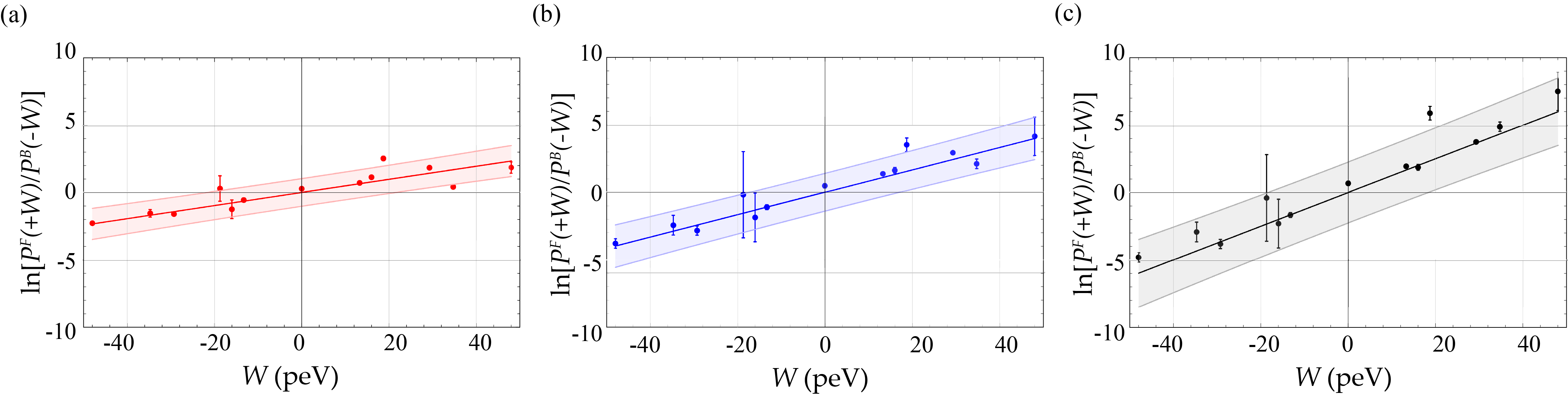}
\par\end{centering}
\caption{{\bf Detailed fluctuation relation for
two interacting spin 1/2. } The experimental results and the linear fit for the logarithm
of the detailed fluctuation relation, \change{$\ln\left[P^{F}\left(+W\right)/P^{B}\left(-W\right)\right]=W\left(k_B T\right)^{-1}$},
are displayed for three different initial-state-preparation temperatures: (a) \change{$k_B T=20\pm3$~$\text{peV}$},  (b) \change{$k_B T=12\pm2$~$\text{peV}$} 
and  (c) \change{$k_B T= 9\pm2$~$\text{peV}$}.
The highlighted  areas between the thin lines in the panels represent the prediction interval
where an observation could fall with confidence level of 99$\%$.
The linear regression is performed excluding these out-layer points: \change{$W = \pm 19.37\text{peV}$} for all panels and also \change{$W=35.92$~$\text{peV}$} for panel (a). The resulting temperatures are: (a) \change{$k_B T = 21\pm2$~$\text{peV}$}, (b)\change{$k_B T=12\pm1$~$\text{peV}$}, and (c) \change{$k_B T=8\pm1$~$\text{peV}$}~\cite{SuppMat}.
}
\label{fig:FIG4}
\end{figure*}

 The obtained statistics  {for $p_{m|n}$
elements} is shown in Fig.~\ref{fig:FIG2}, where the 16 transition
probabilities  {are} identified for both the forward and backward protocols.
  { The experimental transition probabilities obtained from our inversion scheme plus adapted-MLE
method~\citep{SuppMat} and the corresponding theoretical
results obtained by direct numerical simulation of the unitary operations in Fig.~\ref{fig:Foward and Backward circuit} are in very good agreement for both forward and backward protocols. Also, the validity of the micro-reversibility
hypothesis} $p_{m|n}^{F}=p_{n|m}^{B}$ is satisfied
with a good accuracy. 

With 
the method we propose, we can obtain $P(W)$ directly from the  {results in} Fig.~\ref{fig:FIG2} and the measurement of the population
in the initial thermal state.  Experimental results
for the work probability distribution are shown in Fig.~\ref{fig:FIG3}.
 {Transition probabilities between higher energy states are larger for the highest temperature case, with most channels significantly different from zero, than
for the lowest temperature cases, where the population of the ground-state
is higher: compare Fig.~\ref{fig:FIG3}(a)
and (d) with \change{$k_B T \approx 20$~$\text{peV}$}  to Fig.~\ref{fig:FIG3}(c)
and (f), where \change{$k_B T \approx 9 $~$\text{peV}$}}.
\change{To highlight the importance of interactions, we compare in~\cite{SuppMat} the experimental results for the work distribution to its theoretical interacting and non-interacting counterparts.}

 Using  {our} protocol we can also verify
 {the detailed fluctuation relation~\citep{Kurchan2000,Crooks1999,Talkner2007b}
\begin{equation}
\frac{P^{F}\left(+W\right)}{P^{B}\left(-W\right)}=e^{\change{\left(W-\Delta F\right)/\left(k_B T\right)}},
\label{eq:Tasaki-CrookLog}
\end{equation}
 for this interacting spins system
driven out-of-equilibrium.
Here $\Delta F$ is the free energy variation. In our experiment
$\Delta F=0$, since the Hamiltonian after and before the driving
protocol are the same}. {In Fig.~\ref{fig:FIG4}
we show the logarithm of Eq.~\eqref{eq:Tasaki-CrookLog} for the
three different temperatures.} The points represent the experimental
result. The solid line is the linear regression for each spin temperature
data set  { and, from Eq.~\eqref{eq:Tasaki-CrookLog}, the slope of each line gives an estimate of
the corresponding inverse temperature \change{ $\left(k_B T\right)^{-1}$}.} The estimated error
of each point is obtained by means of standard error propagation.
 {The temperatures estimated in this way are in good agreement with
the temperatures of the} initial Gibbs  {states} certified by QST \cite{SuppMat}.

 {\emph{Conclusions.}}  To summarize,
we have proposed and experimentally implemented a new approach to
access energy fluctuations and the work distribution in a  {many}
body  {quantum} system.  {We} have tested our method in an  {  Ising-like}
system  {composed by two spin-1/2 and} driven out-of-equilibrium.   {In addition, by obtaining the bi-stochastic transition probability matrix for the system dynamics at different temperatures, we were able} to verify  {the detailed quantum fluctuation
relation} for  {an} interacting system. The method introduced here can
be applied in  {a} diversity of physical setups to investigate energy fluctuations
and thermodynamical quantities such as, work, heat, and entropy production
in non-equilibrium quantum systems.
\begin{acknowledgments}
\emph{Acknowledgments}. We tank the Multiuser Central Facilities of
UFABC for the technical support. We acknowledge financial support
from UFABC, CNPq, CAPES, FAPESP, and the Royal Society (grant no.
NA140436). J.P.S.P. thanks support from Innovation, Science and Economic
Development Canada, the Government of Ontario, and CIFAR. This research
was performed as part of the Brazilian National Institute of Science
and Technology for Quantum Information (INCT-IQ).
\end{acknowledgments}

\global\long\def\thesection{S-\Roman{section}}
 \setcounter{section}{0} \global\long\def\thefigure{S\arabic{figure}}
 \setcounter{figure}{0} \global\long\def\theequation{S\arabic{equation}}
 \setcounter{equation}{0}\global\long\def\thetable{S\Roman{table}}
 \setcounter{table}{0}
 
\pagebreak

% \newpage

\section*{Supplemental Material}

We provide here supplementary details about the experimental setup, \change{ data analysis, and extensions to large systems.}

\section*{Experimental setup}

For the implementation  of the method and characterization of the energy fluctuations,
we employed a $^{13}$C-labeled CHCl$_{3}$ liquid sample and a 500
MHz Varian NMR spectrometer.The nuclear spins of the $^{1}$H and $^{13}$C atoms of this molecule can be used to represent a two-qubit  system. The Hamiltonian of this system is composed by an interaction term $\mathcal{H}_{int}$
that represents thel interaction between $^{1}$H and $^{13}$C
nuclei  in the molecule and a term $\mathcal{H}_{Z}$
which expresses the interaction with the external magnetic fields~\citep{Oliveira2007}.
 The  interaction term  is mainly due to the scalar coupling
between the Carbon and Hydrogen nuclear spin that reads
\begin{equation}
\mathcal{H}_{int}=\frac{1}{4}hJ\sigma_{z}^{H}\sigma_{z}^{C},\label{eq:interaction}
\end{equation}
where $J\approx215.1$~Hz is the coupling constant. For performing
NMR experiments, the sample is placed inside of a superconducting
magnet where  it is produced a strong static magnetic field $\boldsymbol{B}_{0}\approx11.75~T$. The interaction between the nuclear spins
and this magnetic field aligned along the $z$-axis induces
the Zeeman effect with a Larmor frequency
of $\upsilon_{n}=-\gamma_{n}B_{0}$, where $\gamma_{n}$ is the
gyromagnetic factor characteristic for each nuclear species~\citep{Oliveira2007}. In oder to set correctly the resonance frequency  of both nuclei, the chemical shift should be also considered \citep{Oliveira2007}. Including the chemical shift, the interaction  with the magnetic field can be described by the Hamiltonian
\begin{eqnarray}
\mathcal{H}_{Z} & = & -\frac{1}{2}h\nu_{H}\sigma_{z}^{H}-\frac{1}{2}h\nu_{C}\sigma_{z}^{C},\label{eq:zeeman}
\end{eqnarray}
where $\nu_{H}$ and $\nu_{C}$ are the resonance  frequencies~\citep{Oliveira2007}
of the Hydrogen and Carbon nuclei, respectively. To describe the dynamics of the nuclear
spins, it is used a rotating frame that moves around the $z$-direction
with a traverse frequency offset $\nu_{rf}$, in such frame we obtain
the effective Hamiltonian~\citep{Oliveira2007}

\begin{eqnarray}
\mathcal{H}' & = & -\frac{1}{2}h\left(\nu_{H}-\nu_{rf}\right)\sigma_{z}^{H}-\frac{1}{2}h\left(\nu_{C}-\nu_{rf}\right)\sigma_{z}^{C}\nonumber\\
		&+&\frac{1}{4}hJ\sigma_{z}^{H}\sigma_{z}^{C}.
\end{eqnarray}
\begin{figure}
\begin{centering}
\includegraphics[scale=0.3]{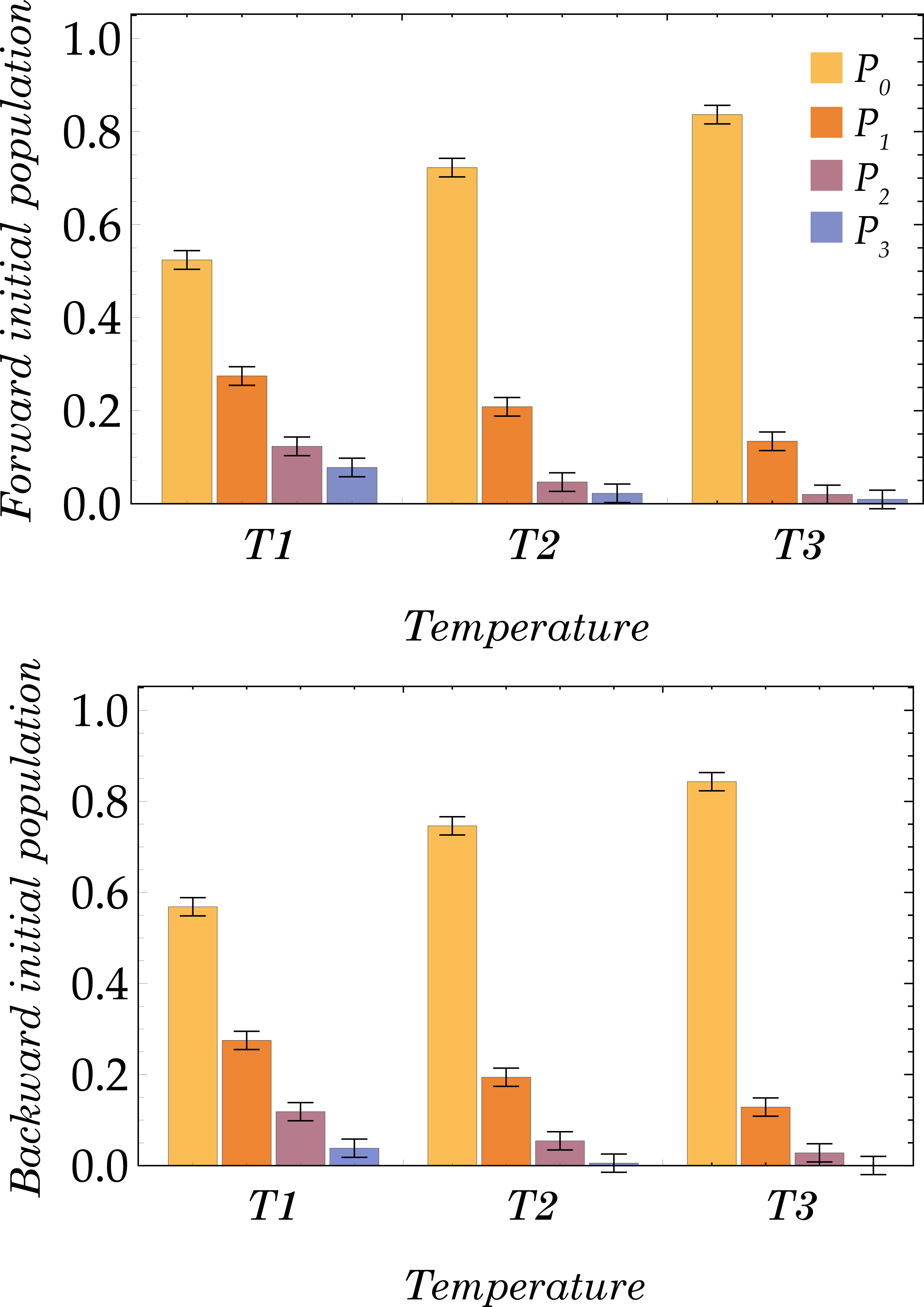}
\par\end{centering}
\centering{}\caption{Populations of the initial state for all effective-spin-temperatures considered.}
\label{FIGS1}
\end{figure}
An important ingredient to perform our experiments is the initialization
of the system. A relevant feature in the NMR experiments is that it
is not possible to deal with a single molecule but with samples containing
an ensemble of many identical molecules, each with the relevant nuclear
spins (qubits). In particular, for the experiment presented here we
have used liquid samples highly diluted in deuterated acetone. This
ensures us that we will have an ensemble of non-interacting molecules,
which means that we have many copies of the same system. For the implementation
of our protocol, we wish to start from thermal states, which are diagonal
in the Hamiltonian basis. Our sample is initially prepared in thermal-like
states corresponding to different spin-temperatures $T$ and such
states can be characterized using a full quantum state tomography~\citep{Oliveira2007a}.

The initial population distribution of the system \change{(corresponding to a thermal Gibbs state)} is shown in Fig.~\ref{FIGS1}
for both the forward and backward protocols. Using these populations
we can estimate the effective spin temperature of the initial pseudo thermal
state by means of the expression \change{$k_B T=\left(E_{2}-E_{0}\right)/\ln\left(\frac{p_{0}}{p_{2}}\right)$},
where $E_{2}-E_{0}$ is the energy difference between the ground state
and the second excited state.These spin temperatures are reported in Table~\ref{TableS1}. 

\noindent 
\begin{table}[h]
\centering{}%
\begin{tabular}{ccccc}
\hline 
 & \multicolumn{4}{c}{\change{$k_B T$ (peV)}}\tabularnewline
 & Forward & Backward & Average & fluctuation relation\tabularnewline
\hline 
\hline 
\change{$k_B T_1$} & 21$\pm$3 & 19$\pm$3 & 20$\pm$3 & 21$\pm$2\tabularnewline
\change{$k_B T_2$} &11$\pm$2 & 12$\pm$2 & 12$\pm$2 & 12$\pm$1\tabularnewline
\change{$k_B T_3$} & 8$\pm$2 & 9$\pm$2 & 9$\pm$2 & 8$\pm$1\tabularnewline
\hline 
\end{tabular}\caption{Characterization of the initial state temperature for the forward
and backward protocol. The effective spin temperature of the prepared
state is estimated from the populations as \change{$k_B T=\left(E_{2}-E_{0}\right)/\ln\left(\frac{p_{0}}{p_{2}}\right)$},
for this experiment \change{$E_{2}-E_{0}=30.3~\text{peV}$}. }
\label{TableS1} 
\end{table}

We implement the forward $\mathcal{U}_{F}$ and backwards $\mathcal{U}_{B}$
evolutions, by applying the unitary operations depicted in the pulse
sequences of Fig. 1 with the  angles display in Table~\ref{TableS2}
\begin{table}[h]
\centering{}%
\begin{tabular}{ccccccc}
\hline 
$i$ & 1 & 2 & 3 & 4 & 5 & 6\tabularnewline
\hline 
\hline 
$\alpha_{i}$ (rad) & 0.48 & -0.80 & $\pi/2$ & -3.61 & 0.69 & $\pi/2$\tabularnewline
$\gamma_{i}$ (rad) & -0.83 & 1.40 & $\pi/2$ & -3.65 & 2.68 & $\pi/2$\tabularnewline
\hline 
\end{tabular}\caption{Parameters used in the pulse sequence displayed in figure~1 on the main text.}
\label{TableS2}
\end{table}

\section*{Magnetization and correlation measurements}

After applying experimentally the protocols that we have described
in the Letter, we measure the magnetization. The obtained values are
reported in Table~\ref{TableS3}. As we can see, the experimental
values are very close to expected theoretical values. However the
influence of experimental imperfections and the uncontrolled coupling
with the environment is the reason we can observe some deviations.
These imperfections affect the statistics of the system and therefore
the transition probabilities: this is why we have to apply the Maximum
Likelihood Estimation (MLE) adapted for such probabilities, as presented
in the following section.\\

\section*{Maximum likelihood estimation to obtain transition probabilities}

For our experiment we measured the magnetization of each nuclear spin
$\langle\sigma_{z}^{H}\rangle$ , $\langle\sigma_{z}^{C}\rangle$
and the correlation function $\langle\sigma_{z}^{H}\sigma_{z}^{C}\rangle$
at three different spin temperatures. Using 
\begin{equation}
\langle\mathcal{O}(\tau)\rangle=\sum_{m,n}\mathcal{O}_{m}p_{n}^{0}p_{m|n}\label{O_tau}
\end{equation}
(see main text) the expression of these mean values are given by

\begin{eqnarray}
\langle\sigma_{z}^{H}\rangle\left(\beta\right) & = & 2\sum_{n}^{N-1}\left(p_{n}^{\beta}-p_{4}^{\beta}\right)p_{1|n}+\sum_{n}^{N-1}\left(p_{n}^{\beta}-p_{4}^{\beta}\right)p_{3|n}\nonumber \\
 &  & +2p_{4}^{\beta}-1,\label{eq:meanvalue1}
\end{eqnarray}

\begin{eqnarray}
\langle\sigma_{z}^{C}\rangle\left(\beta\right) & = & 2\sum_{n}^{N-1}\left(p_{n}^{\beta}-p_{4}^{\beta}\right)p_{1|n}+\sum_{n}^{N-1}\left(p_{n}^{\beta}-p_{4}^{\beta}\right)p_{2|n}\nonumber \\
 &  & +2p_{4}^{\beta}-1,\label{eq:meanvalue2}
\end{eqnarray}

\begin{eqnarray}
\langle\sigma_{z}^{H}\sigma_{z}^{C}\rangle\left(\beta\right) & = & -2\sum_{n}^{N-1}\left(p_{n}^{\beta}-p_{4}^{\beta}\right)p_{2|n}-\sum_{n}^{N-1}\left(p_{n}^{\beta}-p_{4}^{\beta}\right)p_{3|n}\nonumber \\
 &  & -2p_{4}^{\beta}+1.\label{eq:meanvalue3}
\end{eqnarray}

\begin{table}[h]
\centering{}%
\begin{tabular}{ccccc}
\multicolumn{1}{c}{} & \multicolumn{4}{c}{}\tabularnewline
\hline 
 & \multicolumn{4}{c}{\textbf{Effective temperature 1}}\tabularnewline
 & \multicolumn{2}{c}{Experiment} & \multicolumn{2}{c}{Theory}\tabularnewline
\hline 
\hline 
 & Forward & Backward & Forward & Backward\tabularnewline
$\langle\sigma_{z}^{H}\rangle$ & 0.13$\pm$0.05 & -0.14$\pm$0.05 & 0.15 & -0.16\tabularnewline
$\langle\sigma_{z}^{C}\rangle$ & -0.20$\pm$0.05 & 0.07$\pm$0.05 & -0.19 & 0.07\tabularnewline
$\langle\sigma_{z}^{H}\sigma_{z}^{C}\rangle$ & -0.26$\pm$0.05 & 0.13$\pm$0.05 & -0.27 & 0.11\tabularnewline
\hline 
\multicolumn{1}{c}{} & \multicolumn{4}{c}{}\tabularnewline
\hline 
\multicolumn{1}{c}{} & \multicolumn{4}{c}{\textbf{Effective temperature 2}}\tabularnewline
 & \multicolumn{2}{c}{Experiment} & \multicolumn{2}{c}{Theory}\tabularnewline
\hline 
\hline 
 & Forward & Backward & Forward & Backward\tabularnewline
$\langle\sigma_{z}^{H}\rangle$ & 0.26$\pm$0.05 & -0.28$\pm$0.05 & 0.30 & -0.31\tabularnewline
$\langle\sigma_{z}^{C}\rangle$ & -0.32$\pm$0.05 & 0.04$\pm$0.05 & -0.32 & 0.01\tabularnewline
$\langle\sigma_{z}^{H}\sigma_{z}^{C}\rangle$ & -0.39$\pm$0.05 & 0.18$\pm$0.05 & -0.36 & 0.18\tabularnewline
\hline 
\multicolumn{1}{c}{} & \multicolumn{4}{c}{}\tabularnewline
\hline 
\multicolumn{1}{c}{} & \multicolumn{4}{c}{\textbf{Effective spin temperature 3}}\tabularnewline
 & \multicolumn{2}{c}{Experiment} & \multicolumn{2}{c}{Theory}\tabularnewline
\hline 
\hline 
 & Forward & Backward & Forward & Backward\tabularnewline
$\langle\sigma_{z}^{H}\rangle$ & 0.34$\pm$0.05 & -0.38$\pm$0.05 & 0.41 & -0.42\tabularnewline
$\langle\sigma_{z}^{C}\rangle$ & -0.37$\pm$0.05 & 0.01$\pm$0.05 & -0.39 & -0.05\tabularnewline
$\langle\sigma_{z}^{H}\sigma_{z}^{C}\rangle$ & -0.44$\pm$0.05 & 0.17$\pm$0.05 & -0.39 & 0.20\tabularnewline
\hline 
\end{tabular}\caption{Values of the magnetization  and correlation function. }
\label{TableS3} 
\end{table}

Using the expression of the mean values~\eqref{eq:meanvalue1}, \eqref{eq:meanvalue2},
and \eqref{eq:meanvalue3} we construct the matrix equation $\mathcal{A}\boldsymbol{x}=\boldsymbol{b}$
as explained in the main letter (see Eq.~{(}1{)} of the main text).

\begin{widetext}

\begin{equation}
\left[\begin{array}{c}

2\sum_{n}^{N-1}\left(p_{n}^{\beta_{1}}-p_{4}^{\beta_{1}}\right)p_{1|n}+\sum_{n}^{N-1}\left(p_{n}^{\beta_{1}}-p_{4}^{\beta_{1}}\right)p_{3|n}+2p_{4}^{\beta_{1}}-1\\2\sum_{n}^{N-1}\left(p_{n}^{\beta_{1}}-p_{4}^{\beta_{1}}\right)p_{1|n}+\sum_{n}^{N-1}\left(p_{n}^{\beta_{1}}-p_{4}^{\beta_{1}}\right)p_{2|n}+2p_{4}^{\beta_{1}}-1\\-2\sum_{n}^{N-1}\left(p_{n}^{\beta_{1}}-p_{4}^{\beta_{1}}\right)p_{2|n}-\sum_{n}^{N-1}\left(p_{n}^{\beta_{1}}-p_{4}^{\beta_{1}}\right)p_{3|n}-2p_{4}^{\beta_{1}}+1\\2\sum_{n}^{N-1}\left(p_{n}^{\beta_{2}}-p_{4}^{\beta_{2}}\right)p_{1|n}+\sum_{n}^{N-1}\left(p_{n}^{\beta_{2}}-p_{4}^{\beta_{2}}\right)p_{3|n}+2p_{4}^{\beta_{2}}-1\\2\sum_{n}^{N-1}\left(p_{n}^{\beta_{2}}-p_{4}^{\beta_{2}}\right)p_{1|n}+\sum_{n}^{N-1}\left(p_{n}^{\beta_{2}}-p_{4}^{\beta_{2}}\right)p_{2|n}+2p_{4}^{\beta_{2}}-1\\-2\sum_{n}^{N-1}\left(p_{n}^{\beta_{2}}-p_{4}^{\beta_{2}}\right)p_{2|n}-\sum_{n}^{N-1}\left(p_{n}^{\beta_{2}}-p_{4}^{\beta_{2}}\right)p_{3|n}-2p_{4}^{\beta_{2}}+1\\2\sum_{n}^{N-1}\left(p_{n}^{\beta_{3}}-p_{4}^{\beta_{3}}\right)p_{1|n}+\sum_{n}^{N-1}\left(p_{n}^{\beta_{3}}-p_{4}^{\beta_{3}}\right)p_{3|n}+2p_{4}^{\beta_{3}}-1\\2\sum_{n}^{N-1}\left(p_{n}^{\beta_{3}}-p_{4}^{\beta_{3}}\right)p_{1|n}+\sum_{n}^{N-1}\left(p_{n}^{\beta_{3}}-p_{4}^{\beta_{3}}\right)p_{2|n}+2p_{4}^{\beta_{3}}-1\\-2\sum_{n}^{N-1}\left(p_{n}^{\beta_{3}}-p_{4}^{\beta_{3}}\right)p_{2|n}-\sum_{n}^{N-1}\left(p_{n}^{\beta_{3}}-p_{4}^{\beta_{3}}\right)p_{3|n}-2p_{4}^{\beta_{3}}+1\end{array}\right]
=\left[\begin{array}{c}\langle\sigma_{z}^{H}\rangle\left(\beta_{1}\right)\\ \langle\sigma_{z}^{C}\rangle\left(\beta_{1}\right)\\ \langle\sigma_{z}^{H}\sigma_{z}^{C}\rangle\left(\beta_{1}\right)\\ \langle\sigma_{z}^{H}\rangle\left(\beta_{2}\right)\\ \langle\sigma_{z}^{C}\rangle\left(\beta_{2}\right)\\ \langle\sigma_{z}^{H}\sigma_{z}^{C}\rangle\left(\beta_{2}\right)\\ \langle\sigma_{z}^{H}\rangle\left(\beta_{3}\right)\\ \langle\sigma_{z}^{C}\rangle\left(\beta_{3}\right)\\ \langle\sigma_{z}^{H}\sigma_{z}^{C}\rangle\left(\beta_{3}\right)

\end{array}\right]
\label{eq:general1}
\end{equation}
\end{widetext}

It is important to stress that if the measured data were perfect (without
any noise), as in a numerical simulation, the method leads directly
to the correct result. However due to experimental noise, the direct
solution of the Eq.~\textbf{\eqref{eq:general1}} from the experimental
data could yield to nonphysical results. We can overcome this problem
by searching in all the possible valid solutions (in the sense of
the set of transition probabilities obtained being a valid bistochastic
matrix) and extract the closest physical probabilities that solve
Eqs.~\textbf{\eqref{eq:general1}}. Those physical probabilities
in general will not reproduce exactly the measured data, but they
will be the most likely to produce the experimental observation. A
similar problem occurs in quantum state tomography. Here, the usual
procedure to apply the MLE is to use a particular representation for
the density matrix given by 
\begin{eqnarray}
\rho & = & \frac{\mathcal{T}\mathcal{T}^{\dagger}}{\text{Tr}\left(\mathcal{T}\mathcal{T}^{\dagger}\right)},\label{eq:densitydecomp}
\end{eqnarray}
where $\mathcal{T}$ is a triangular matrix. Equation~\eqref{eq:densitydecomp}
guarantees that the density matrix will be a Hermitian defined positive
matrix with trace one. For example, considering a system with two
qubits, we can write $\mathcal{T}$ as 
\begin{equation}
\mathcal{T}=\left(\begin{array}{cccc}
t_{1} & 0 & 0 & 0\\
t_{5}+it_{6} & t_{2} & 0 & 0\\
t_{7}+it_{8} & t_{9}+it_{10} & t_{3} & 0\\
t_{11}+it_{12} & t_{13}+it_{14} & t_{15}+it_{16} & t_{4}
\end{array}\right),
\end{equation}
where $t_{i}$ are the parameters that have to be defined in the optimization
of the maximum likelihood function given the experimental data set.

However, for a transition probability matrix, things are more complicated.
For a transition probability matrix $P$, one could think to resort
to the Birkhoff--von Neumann decomposition~\citep{Dufosse2016}.
This states that for a bistochastic matrix $P$ -- i.e. $\sum_{m}P_{m|n}=\sum_{n}P_{m|n}=1$
-- there exist the parameters $\theta_{1}$, $\theta_{2}$,$\ldots,$$\theta_{k}$~$\in$~$\left(0,1\right)$
with $\sum_{j=}^{k}\theta_{j}=1$ and the permutation matrices $\Lambda_{1}$,
$\Lambda_{2}$,$\ldots,$$\Lambda_{k}$ such that $P$ can be decomposed
as 
\begin{equation}
P=\theta_{1}\Lambda_{1}+\theta_{2}\Lambda_{2}+\ldots+\theta_{k}\Lambda_{k}.
\end{equation}
A permutation matrix $\Lambda_{j}$ is a square matrix whose rows
and columns contain exactly one nonzero entry, which is 1. However
this representation may, in general, not be unique, and finding the
representation with the minimum number of terms has been shown to
be NP-hard problem~\citep{Dufosse2016}. Due to the difficulty of
using such a representation we are going to use a different approach
and instead of defining a specific representation for the transition
probability matrix, we will solve Eq.~\textbf{\eqref{eq:general1}}.
From that solution and by means of the numerical optimizations explained
below, we will obtain the closest physical solution.

Based on this, we define our likelihood function as 
\begin{equation}
\mathcal{F}\left(x\right)=\sum_{k,j}\left(x_{k,j}-\varXi_{k,j}\right)^{2},\label{eq:likelihood}
\end{equation}
where $x_{k,j}$ are the solution of the system of equations given
in~\textbf{\eqref{eq:general1}} as constructed from the set of the
experimentally measured observables, and $\varXi$ is a positive matrix,
whose elements satisfy the condition $0\leq \varXi_{k,j}\leq1$. Then, we
find the minimum of Eq.~\eqref{eq:likelihood} to obtain the closest
positive matrix $\varXi$ that better fit our expected result. Doing
the minimization we only guarantee that the elements of the optimized
matrix are probabilities, but such elements also have to satisfy the
constraint $\sum_{k}\varXi_{k,j}=\sum_{j}\varXi_{k,j}=1$. We can
transform the positive matrix $\varXi$ into a bistochastic one using
the Sinkhorn-Knopp algorithm~\citep{Sinkhorn1967}. That algorithm
is a simple iterative method that generates a bistochastic matrix
by alternatively normalizing the rows and the columns of a positive
matrix. After getting the bistochastic matrix by the application of
the Sinkhorn-Knopp algorithm, we repeat the minimization by using
the likelihood function~\eqref{eq:likelihood} where this solution
enters in the next cycle as a new $\varXi_{k,j}$ and start the process over again until finally
we reach a convergence in the final solution. For our calculation
we are setting a threshold of 1000 iteration with a tolerance of $1\times10^{-6}$, but the convergence of this protocol is fast, for the forward protocol
we needed 3 iterations and for the backward protocol 13 iterations
only.

\change{
\section*{Systems larger than few bodies}

\subsection{Initial state}
The reconstruction of the system's initial quantum state  is not strictly necessary at the beginning of the protocol.
Quantum state tomography (QST) for the initial state can in fact be avoided as long as it can be reasonably assumed that the initial state is known, e.g. because the system has been carefully prepared in a desired state or has been allowed to thermalize at a known temperature. 

In addition, for large systems, there are by now several methods which significantly improve on the scaling of QST with system size,  making its scaling less than exponential or providing a good approximation for the state. Possibilities are, for instance, compressed sensing QST~\cite{compressed_sensing}, quantum state learning \cite{quantum_state_learning}, QST assisted by machine learning \cite{machine_learning}. 

\subsection{Adapting the method to large systems}

\subsubsection{Systems  described by a lattice Hamiltonian}

For lattice systems, the operators to be measured to derive the work distribution function should be chosen based on what is easy to measure for the specific system at hand. However, for systems bigger than few-bodies, the number of transitions would become prohibitively large to be reconstructed one-by-one, by this -- and possibly any other  -- method. 
In this case it is suggested to use an interpolation approach within the proposed method to reconstruct the work distribution.

For large many-body systems, the number of transitions increases rapidly with size:
we expect then the work probability distribution to regularise very rapidly to a fairly well-defined bell shape. Indeed, we observed this already for systems of the order of 6-8 spins \cite{Krissia}. In this case, it would be superfluous to measure this curve for every single transition, while a coarse-grained sampling would be sufficient to reconstruct the curve. The solution of Eq. (1) in our paper could then be adapted to the case in which a simple continuous function is assumed for the distribution. The same approximated methods for QST for large systems mentioned in the previous subsection could be adapted to solve this problem. 

An interesting case would be the one of very strongly correlated systems. In this case correlations reduce -- de facto -- the Hilbert space to very few allowed transitions, as seen for example in Fig. 1 (b) and (d) in Ref. \cite{Krissia}.  Here the original scheme as currently described in the paper could be used.
}

\change{
\subsubsection{Continuous, interacting $N$-particle systems}

Within the context of density functional theory (DFT), Ref.~\cite{Rohringer2006} proposed two functionals for the transition matrix $p_{m|n}$ and discussed their limitations. These functionals are derived via an inversion scheme, and their key ingredient is the time-dependent electronic local density. More recently, Ref.~\cite{Beau2020} proposed a way to access energy fluctuations in scale-invariant, continuous  quantum fluids.
In the following, we highlight how our method can be adapted to systems of continuous variables.

For continuous, interacting $N$-particle systems, the measure of a single local operator which commutes with the final Hamiltonian $H_\tau$  is in principle sufficient to determine the transition matrix $p_{m|n}$ and hence the work probability distribution. 

In this case we can write Eq.~(\ref{O_tau}) as
\begin{eqnarray}
\mathcal{O}({\bf{r}},
\tau)&=& \text{Tr} [\hat{O}({\bf{r}})\hat{\rho}(t)]    \\
&=&\sum_{m,n}^{M}\mathcal{O}({\bf{r}})_{m}p_{n}^{0}p_{m|n},\label{O_r_tau}
\end{eqnarray}
where both  $\mathcal{O}({\bf{r}},\tau)$, and $\mathcal{O}({\bf{r}})_{m}$
 are local and measurable. For example, where appropriate,  $\mathcal{O}({\bf{r}},
\tau)$ could be the local particle density $n({\bf{r}},
\tau)$ or the local magnetization $m({\bf{r}},
\tau)$. For electronic systems, the advantage of these observables is that good estimates of them could be calculated using DFT, which could facilitate the inversion scheme.

Because the system is continuous, there is, in principle, no limit to the number of points  ${\bf{r}}$ where these measurements could be taken, and hence the measurement of a single local operator is enough to recover all elements of the matrix $p_{m|n}$.
A more practical approach would be to perform measurements over a manageable set of points $({\bf{r}}_1, {\bf{r}}_2,\dots {\bf{r}}_s)$ and then fit the sets of measurements to continuous functions.
In addition, for electronic systems, DFT methods~\cite{Rohringer2006a} could be used to get approximations for $p_{m|n}$, which could aid the inversion scheme.

In electronic systems, the number of eigenstates $M$ is generally very large, for all practical purposes infinite. This could be tackle by  truncating the basis appropriately with respect to the characteristic energies (including the spin temperature) of the system at hand.

We aim to explore these points further in  future developments of the method. 
}

\begin{figure*}
\begin{centering}
\includegraphics[scale=0.25]{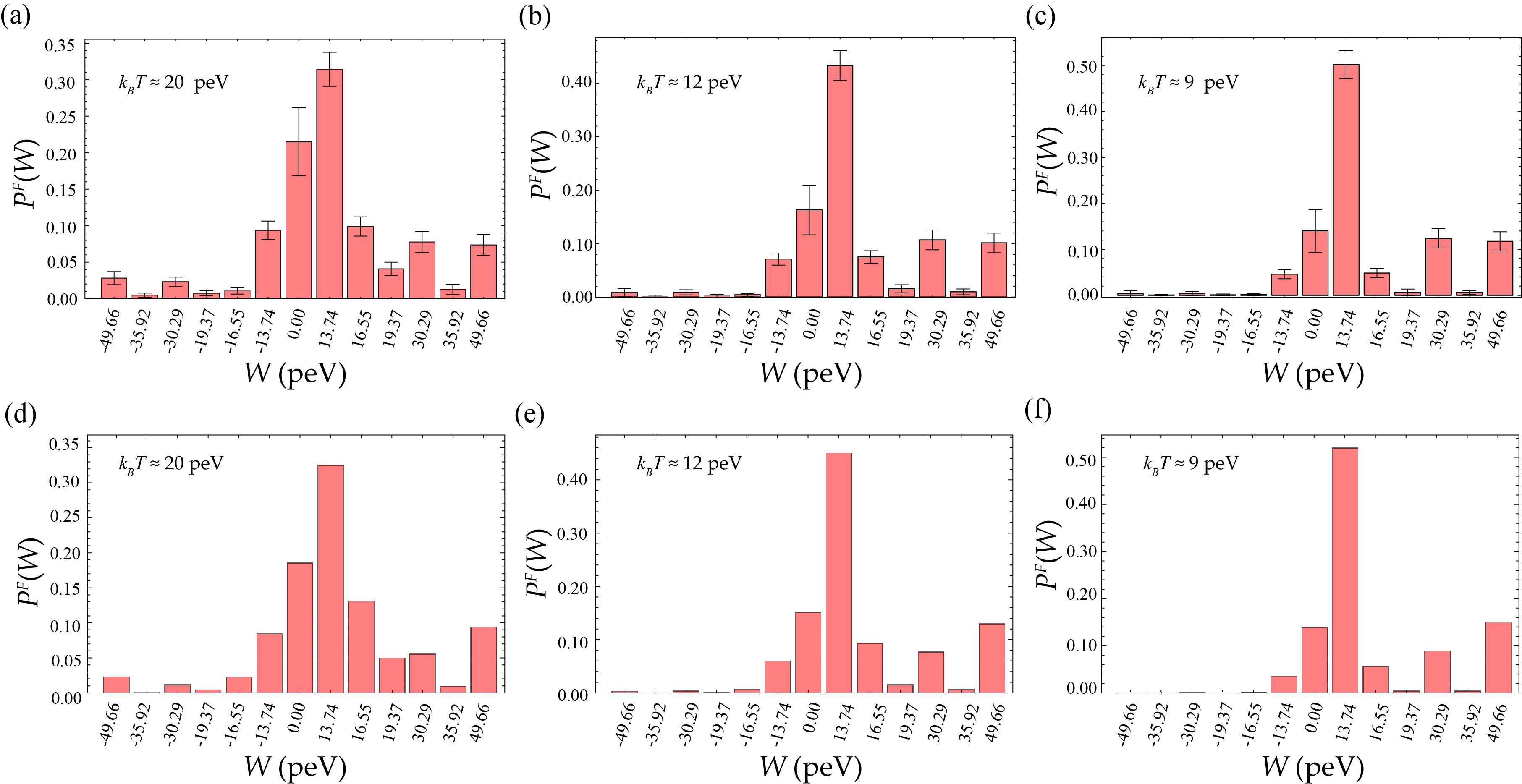}
\par\end{centering}
\caption{\change{Experimental (a, b, c) and theoretical (d, e, f) results for the forward work distribution for the three different initial spin temperatures (a, d) \change{$k_B T = 20\pm3$~$\text{peV}$},  (b, e) \change{$k_B T= 12\pm2$~$\text{peV}$} 
and  (c, f) \change{$k_B T = 9\pm2$~$\text{peV}$}}}.
\label{fig:forw}
\end{figure*}

\begin{figure*}
\begin{centering}
\includegraphics[scale=0.25]{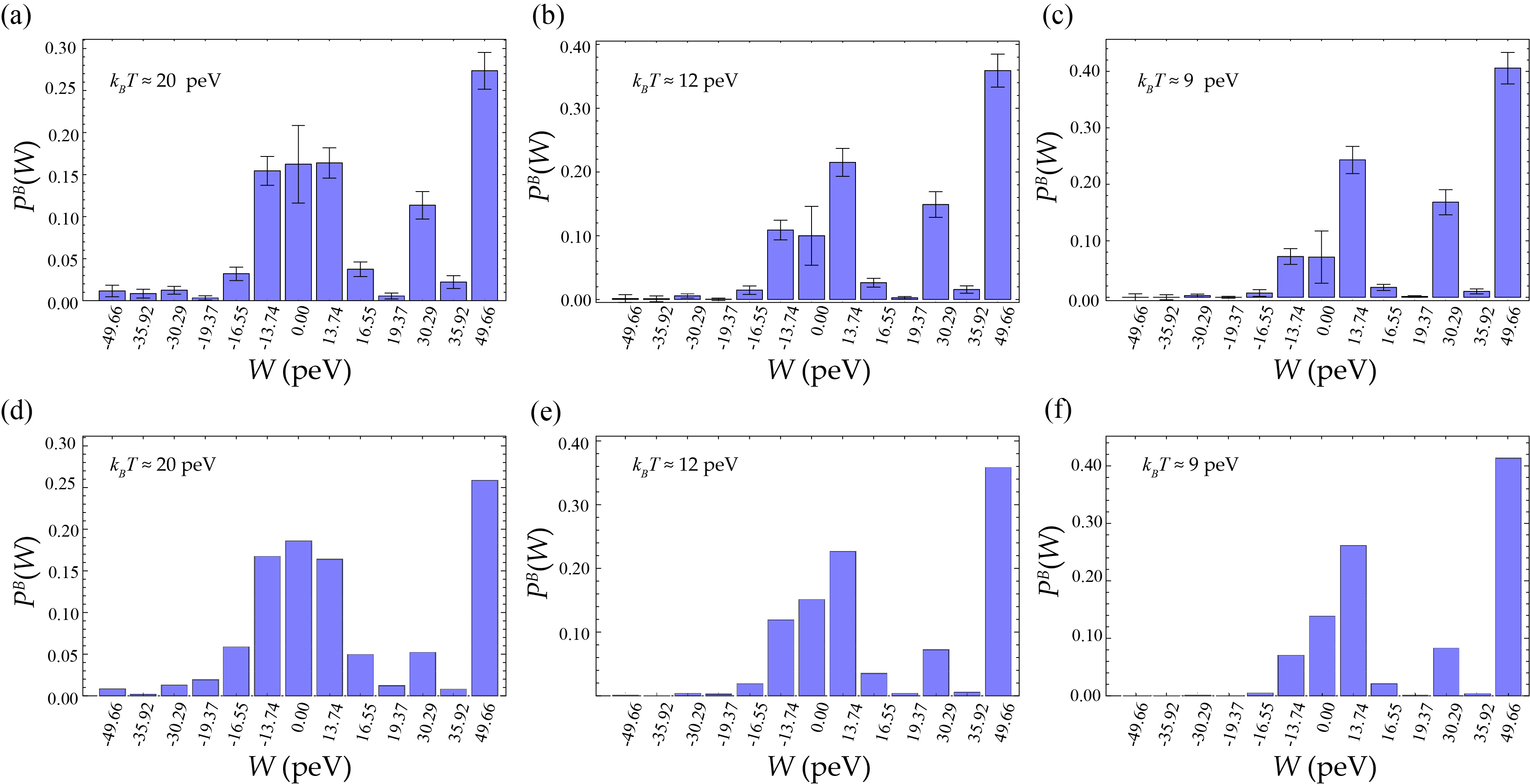}
\par\end{centering}
\caption{\change{Experimental (a, b, c) and theoretical (d, e, f) results for the backward work distribution for the three different initial spin temperatures (a, d) \change{$k_B T = 20\pm3$~$\text{peV}$},  (b, e) \change{$k_B T= 12\pm2$~$\text{peV}$} 
and  (c, f) \change{$k_B T = 9\pm2$~$\text{peV}$}}.
 }
\label{fig:back}
\end{figure*}

\change{
\section*{Comparison between experimental results and theoretical simulations for the forward and backward stroke work distribution}

 We are reporting in this  section the comparison between the experimentally extracted work distributions in Fig.  3 of the main text with the  corresponding theoretical  simulations, for both the forward (Fig.~\ref{fig:forw}) and the backward (Fig.~\ref{fig:back}) process. The technical imprecision is associated to the errors bars in the experimental data. Indeed, these figures confirms the very good agreement between the theoretical predictions and results from the experiment, with the great majority of experimentally extracted work probabilities being well within one error bar of the predicted results.
 }

\begin{figure*}
\begin{centering}
\includegraphics[scale=0.25]{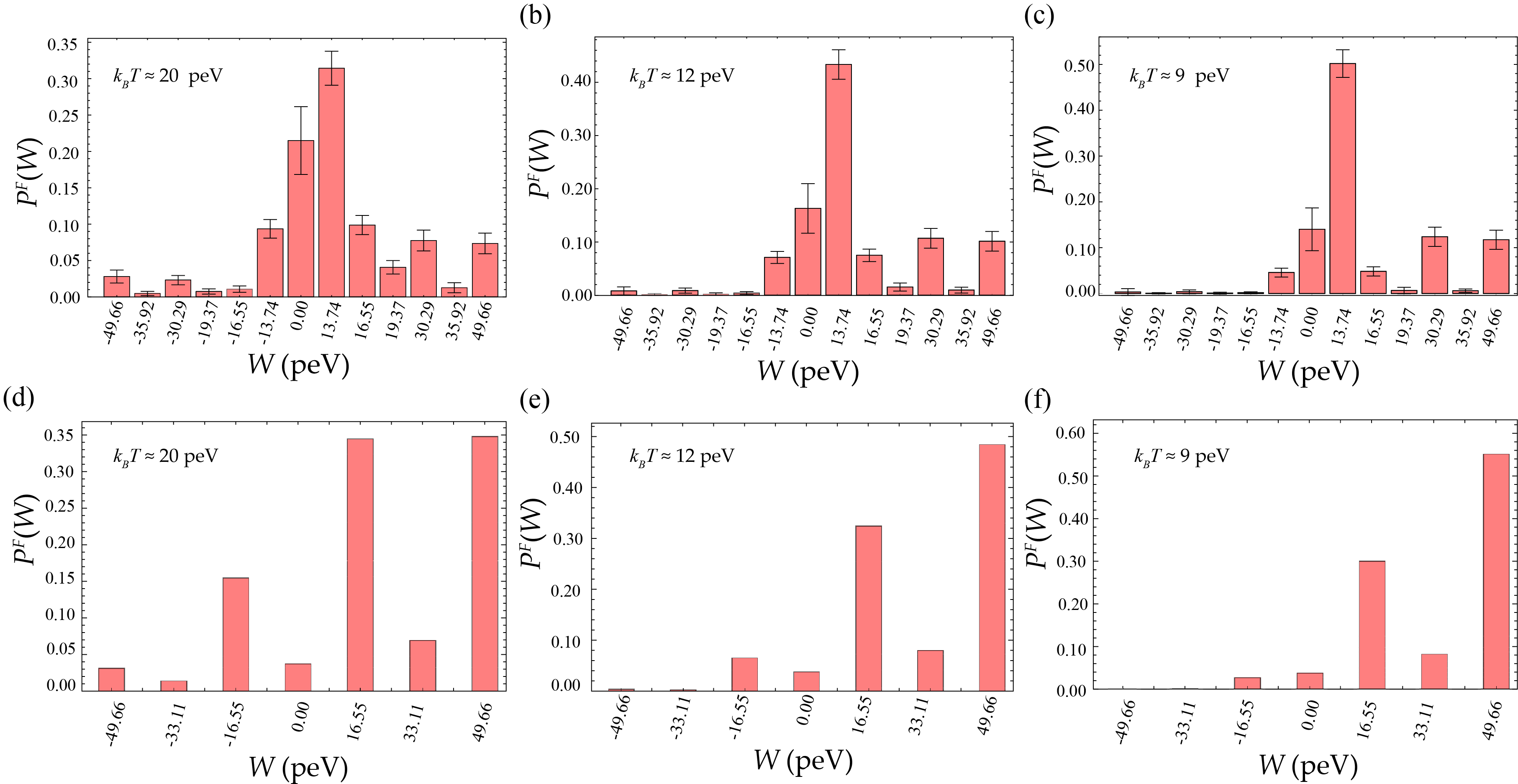}
\par\end{centering}
\caption{\change{Experimental (a, b, c) and non-interacting theoretical (d, e, f) results for the forward work distribution for the three different initial spin temperatures (a, d) \change{$k_B T = 20\pm3$~$\text{peV}$},  (b, e) \change{$k_B T= 12\pm2$~$\text{peV}$} 
and  (c, f) \change{$k_B T = 9\pm2$~$\text{peV}$}}.
 }
\label{fig:forw_ni}
\end{figure*}

\begin{figure*}
\begin{centering}
\includegraphics[scale=0.25]{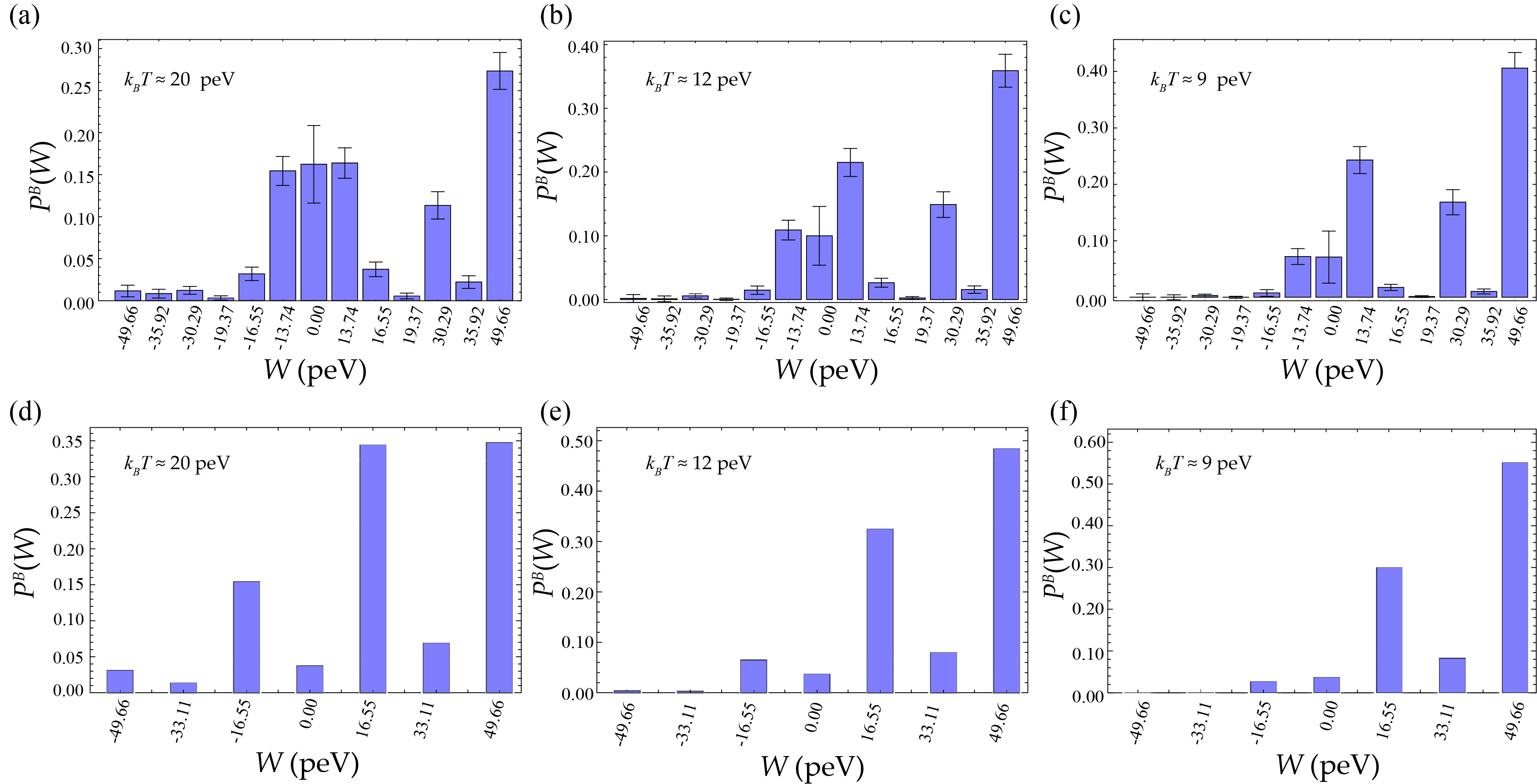}
\par\end{centering}
\caption{\change{Experimental (a, b, c) and non-interacting theoretical  (d, e, f) results for the backward work distribution for the three different initial spin temperatures (a, d) \change{$k_B T = 20\pm3$~$\text{peV}$},  (b, e) \change{$k_B T= 12\pm2$~$\text{peV}$} 
and  (c, f) \change{$k_B T = 9\pm2$~$\text{peV}$}}.
 }
\label{fig:back_ni}
\end{figure*}

\change{
\section*{Comparison between experimental results and the corresponding non-interacting theoretical system for the forward and backward stroke work distribution}

 We are reporting in this  section the comparison between the experimentally extracted work distributions in Fig.  3 of the main text and the simulations' results for the corresponding non-interacting system ($J=0$), for both the forward (Fig.~\ref{fig:forw_ni}) and the backward (Fig.~\ref{fig:back_ni}) process. As can be observed the non-interacting results are qualitatively and quantitatively distinct from the experimental ones, underlying the importance of  interactions at the temperatures considered.
 }

{1}


\begin{thebibliography}{10}
\bibitem{Goold2016}J. Goold, M. Huber, A. Riera, L. del Rio, and
P. Skrzypczyk, The role of quantum information in thermodynamics --
a topical review, J. Phys. A: Math. Theor. \textbf{49}, 143001 (2016).

\bibitem{Vinjanampathy2016}S. Vinjanampathy and J. Anders, Quantum
thermodynamics, Contemp. Phys. \textbf{57}, 545 (2016).

\bibitem{Kosloff2013}R. Kosloff, Quantum Thermodynamics: A Dynamic
Viewpoint, Entropy \textbf{15}, 2100 (2013).

\bibitem{Jarzynski1997}C. Jarzynski, Nonequilibrium Equality for
Free Energy Differences, Phys. Rev. Lett. \textbf{78}, 2690 (1997).

\bibitem{Crooks1999}G. E. Crooks Entropy Production Fluctuation Theorem
and the Nonequilibrium Work Relation for Free-Energy Differences,
Phys. Rev. E \textbf{60}, 2721 (1999).

\bibitem{Kurchan2000}J. Kurchan, A quantum fluctuation theorem, cond-mat/0007360;
H. Tasaki, and C. Jarzynski, Relations for Quantum Systems and Some
Applications, cond-mat/0009244v2.

\bibitem{Mukamel2003}S. Mukamel, Quantum extension of the Jarzynski
relation: Analogy with stochastic dephasing, Phys. Rev. Lett. \textbf{90},
170604 (2003).

\bibitem{Talkner2007b}P. Talkner, and P. Hänggi, The Tasaki-Crooks
quantum fluctuation theorem, J. Phys. A \textbf{40}, F569 (2007).

\bibitem{Esposito2009} M. Esposito, U. Harbola, and S. Mukamel, non-equilibrium
fluctuations, fluctuation theorems, and counting statistics in quantum
systems, Rev. Mod. Phys. \textbf{81}, 1665 (2009).

\bibitem{Campisi2011}M. Campisi, P. Hänggi, and P. Talkner, Colloquium:
Quantum fluctuation relations: Foundations and applications, Rev.
Mod. Phys. \textbf{83}, 771 (2011).

\bibitem{Hanggi2015} P. Hänggi and P. Talkner, The other QFT, Nat.
Phys. \textbf{11}, 108 (2015).

\bibitem{Camati2018}P. A. Camati and R. M. Serra, Verifying detailed
fluctuation relations for discrete feedback-controlled quantum dynamics,
Phys. Rev. A \textbf{97}, 042127 (2018).

\bibitem{Dorner2013}R. Dorner, S.R. Clark, L. Heaney, R. Fazio, J.
Goold, and V. Vedral, Extracting quantum work statistics and fluctuation
theorems by single-qubit interferometry, Phys. Rev. Lett. \textbf{110},
230601 (2013).

\bibitem{Mazzola2013}L. Mazzola, G. De Chiara, and M. Paternostro,
Measuring the characteristic function of the work distribution, Phys.
Rev. Lett. \textbf{110}, 230602 (2013).

\bibitem{Batalhao2014}T. B. Batalhão, A. M. Souza, L. Mazzola, R.
Auccaise, R. S. Sarthour, I. S. Oliveira, J. Goold, G. De Chiara,
M. Paternostro, and R. M. Serra, Experimental Reconstruction of Work
Distribution and Study of Fluctuation Relations in a Closed Quantum
System, \textit{\emph{Phys. Rev. Lett.}} \textbf{113}, 140601 (2014).

\bibitem{Batalhao2015}T. B. Batalhão, A.\LyXThinSpace M. Souza, R.\LyXThinSpace S.
Sarthour, I.\LyXThinSpace S. Oliveira, M. Paternostro, E. Lutz, R.\LyXThinSpace M.
Serra, Irreversibility and the Arrow of Time in a Quenched Quantum
System. Phys. Rev. Lett. \textbf{115}, 190601 (2015).

\bibitem{Camati2016}P. A. Camati, J. P.\LyXThinSpace S. Peterson,
T. B. Batalhão, K. Micadei, A. M. Souza, R. S. Sarthour, I. S. Oliveira,
and R. M. Serra, Experimental Rectification of Entropy Production
by Maxwell's Demon in a Quantum System, \textit{\emph{Phys. Rev. Lett.}}\emph{
}\textbf{117}, 240502 (2016).

\bibitem{Peterson2018}J. P. S. Peterson, T. B. Batalhão, M. Herrera,
A. M. Souza, R. S. Sarthour, I. S. Oliveira, R. M. Serra, Experimental
characterization of a spin quantum heat engine, Phys. Rev. Lett. \textbf{123},
240601 (2019).

\bibitem{Cerisola2017}F. Cerisola, Y. Margalit, S. Machluf, A. J.
Roncaglia, J. P. Paz and R. Folman , Using a quantum work meter to
test non-equilibrium fluctuation theorems, Nat. Commun. \textbf{8},
1241 (2017).

\bibitem{Jones2015}R. O. Jones, Density functional theory: Its origins, rise to prominence, and future, Rev. Mod. Phys. 87, 897 (2015).

\bibitem{Herrera2017}M. Herrera, R. M. Serra, and I. D’Amico, Scientific
reports \textbf{7}, 4655 (2017)

\bibitem{Herrera2018} M. Herrera, K. Zawadzki, and I. D'Amico, Eur. Phys. J. B \textbf{91}, 248 (2018)

\bibitem{Skelt2019} A. H. Skelt, K. Zawadzki and I. D'Amico, J. Phys. A: Math. Theor. 52, 485304 (2019)

\bibitem{Talkner2007}P. Talkner, E. Lutz, and P. Hänggi, Fluctuation
theorems: Work is not an observable, \textit{\emph{Phys. Rev. E}}
\textbf{75}, 050102(R) (2007).

\bibitem{Talkner2016} P. Talkner and P. Hänggi, Aspects of quantum
work, Phys. Rev. E \textbf{93}, 022131 (2016).

\bibitem{Seifert2012}U. Seifert, Stochastic thermodynamics, fluctuation
theorems and molecular machines, Rep. Prog. Phys. \textbf{75}, 126001
(2012).

\bibitem{Suomela2014}S. Suomela, P. Solinas, J. P. Pekola, J. Ankerhold,
and T. Ala-Nissila, Phys. Rev. B \textbf{90}, 094304 (2014).

\bibitem{Monnai2005}T. Monnai, Unified treatment of the quantum fluctuation
theorem and the Jarzynski equality in terms of microscopic reversibility,
Phys. Rev. E \textbf{72}, 027102 (2005).

\bibitem{Watanabe2014}G. Watanabe, B. P. Venkatesh, and P. Talkner,
Generalized energy measurements and modified transient quantum fluctuation
theorems, Phys. Rev. E \textbf{89}, 052116 (2014).

\bibitem{Roncaglia2014}A. J. Roncaglia, F. Cerisola, and J. P. Paz,
Work measurement as a generalized quantum measurement, Phys. Rev.
Lett. \textbf{113}, 250601 (2014).

\bibitem{Rohringer2006}N. Rohringer, S. Peter, and J. Burgdorfer,
Calculating state-to-state transition probabilities within time-dependent
density-functional theory, Phys. Rev. A \textbf{74}, 042512 (2006).

\bibitem{SuppMat}See Supplemental Material for more details.

\bibitem{An2014}S. An, J.-N. Zhang, M. Um, D. Lv, Y. Lu, J. Zhang,
Z.-Q. Yin, H. Quan, and K. Kim, Nat. Phys. \textbf{11}, 193 (2014).

\bibitem{Peterson2016}J. P. Peterson, R. S. Sarthour, A. M. Souza,
I. S. Oliveira, J. Goold, K. Modi, D. O. Soares-Pinto, and L. C. Celeri,
Proc. R. Soc. A \textbf{472}, 20150813 (2016).

\bibitem{Oliveira2007}I. S. Oliveira, T. J. Bonagamba, R. S. Sarthour,
J. C. C. Freitas, and R. R. deAzevedo, NMR Quantum Information Processing
(Elsevier, Amsterdam, 2007).

\bibitem{Vatan2004}F. Vatan and C. Williams, Optimal quantum circuits
for general two-qubit gates, Phys. Rev. A \textbf{69}, 032315 (2004).

\bibitem{Chuang1997}I. L. Chuang and M. A. Nielsen, Prescription
for experimental determination of the dynamics of a quantum black
box, J. Mod. Opt. \textbf{44}, 2455 (1997).

\bibitem{Nielsen2011}M. A. Nielsen and I. L. Chuang, \emph{Quantum
Computation and Quantum Information} (Cambridge University Press,
2011).

\bibitem{Banaszek1999}K. Banaszek, G. M. A. D'Ariano, M. G. Paris, and M. F. Sacchi, Maximum-likelihood estimation of the density matrix, Phys.
Rev. A \textbf{61}, 010304 (1999).

\bibitem{James2001}D. F. V. James, P. G. Kwiat, W. J. Munro, and A.
G. White, Measurement of qubits, Phys. Rev. A \textbf{64}, 052312
(2001).

\bibitem{Micadei2019}K. Micadei, J. P. S. Peterson, A. M. Souza,
R. S. Sarthour, I. S. Oliveira, G. T. Landi, T. B. Batalhão, R. M.
Serra, and E. Lutz, Reversing the direction of heat flow using quantum
correlations, Nat. Commun. \textbf{10}, 2456 (2019).


\bibitem{note1}{\change{We note that instead the strength of many-body interactions {\it within} the system does not restrict the protocol speed.}}

\bibitem{note2}{\change{In principle one could apply the method even if $\mathcal{O}$ and $\mathcal{H}_{\tau}$ would not commute, however this
would require many more measurements.  Depending on the specific experimental scenario, our method can be adapted according to the
available observables and range of temperatures.}}

\end{thebibliography}

\begin{thebibliography}{1}
\bibitem{Oliveira2007a}I. S. Oliveira, T. J. Bonagamba, R. S. Sarthour,
J. C. C. Freitas, and R. R. deAzevedo, NMR Quantum Information Processing
(Elsevier, Amsterdam, 2007) 

\bibitem{Dufosse2016}F. Dufossé and B. Uçar, Notes on Birkhoff-von
Neumann decomposition of doubly stochastic matrices,
Linear Algebra Appl. \textbf{497}, 108 (2016).

\bibitem{Sinkhorn1967}R. Sinkhorn and P. Knopp, Concerning nonnegative
matrices and doubly stochastic matrices, Pacic J. Math.
\textbf{21}, 343 (1967). 

\bibitem{Rohringer2006a}N. Rohringer, S. Peter, and J. Burgdorfer,
Calculating state-to-state transition probabilities within time-dependent
density-functional theory, Phys. Rev. A \textbf{74}, 042512 (2006).

\bibitem{compressed_sensing} Anastasios Kyrillidis, Amir Kalev, Dohyung Park, Srinadh Bhojanapalli, Constantine Caramanis,  Sujay Sanghavi, Provable compressed sensing quantum state tomography via non-convex methods, NPJ Quantum Information \textbf{4}, 36 (2018); 
A Steffens, C A Riofrío, W McCutcheon, I Roth, B A Bell, A McMillan, M S Tame, J G Rarity and J Eisert, Experimentally exploring compressed sensing quantum tomography, Quantum Sci. Technol. \textbf{2}, 025005 (2017). \bibitem{quantum_state_learning} A. Rocchetto, S. Aaronson, S. Severini, G. Carvacho, D. Poderini, I. Agresti, M. Bentivegna, and F. Sciarrino, Experimental learning of quantum states, Sci. Adv. \textbf{5}, eaau1946 (2019).
\bibitem{machine_learning} Sanjaya Lohani, Brian T Kirby, Michael Brodsky, Onur Danaci and Ryan T Glasser, Machine learning assisted quantum state estimation, Mach. Learn.: Sci. Technol. \textbf{1}, 035007 (2020)

\bibitem{Krissia} Krissia Zawadzki, Roberto M. Serra, and Irene D'Amico, Work-distribution quantumness and irreversibility when crossing a quantum phase transition in finite time, Phys. Rev. Research \textbf{2}, 033167 (2020).

\bibitem{Beau2020} Mathieu Beau and Adolfo del Campo, Nonadiabatic Energy Fluctuations of Scale-Invariant Quantum Systems in a Time-Dependent Trap, Entropy \textbf{22}, 515 (2020).

\end{thebibliography}
\end{document}